\documentclass[journal]{IEEEtran}
\usepackage{amsmath,amsfonts}
\usepackage{algorithmic}
\usepackage{algorithm}
\usepackage{array}
\usepackage[caption=false,font=normalsize,labelfont=sf,textfont=sf]{subfig}
\usepackage{textcomp}
\usepackage{stfloats}
\usepackage{url}
\usepackage{verbatim}
\usepackage{graphicx}
\usepackage{cite}
\begin{document}

\title{A 3D Mobile Crowdsensing Framework\\for Sustainable Urban Digital Twins}

\author{
  Taku Yamazaki,
  Kaito Watanabe,
  Tatsuya Kase,
  Kenta Hasegawa,
  Koki Saida,
  and Takumi Miyoshi
  \thanks{T. Yamazaki, K. Watanabe, T. Kase, K. Hasegawa, K. Saida, and T. Miyoshi are with the Shibaura Institute of Technology, Saitama-shi, Saitama, 337-8570 Japan e-mail: (taku@shibaura-it.ac.jp).}
\thanks{This work has been submitted to the IEEE for possible publication. Copyright may be transferred without notice, after which this version may no longer be accessible.}
}

\markboth{A 3D Mobile Crowdsensing Framework for Sustainable Urban Digital Twins}%
{A 3D Mobile Crowdsensing Framework for Sustainable Urban Digital Twins}

\maketitle

\begin{abstract}
In this article, we propose a 3D mobile crowdsensing (3D-MCS) framework aimed at sustainable urban digital twins (UDTs). The framework comprises four key mechanisms: (1) the 3D-MCS mechanism, consisting of active and passive models; (2) the Geohash-based spatial information management mechanism; (3) the dynamic point cloud integration mechanism for UDTs; and (4) the web-based real-time visualizer for 3D-MCS and UDTs. The active sensing model features a gamified 3D-MCS approach, where participants collect point cloud data through an augmented reality territory coloring game. In contrast, the passive sensing model employs a wearable 3D-MCS approach, where participants wear smartphones around their necks without disrupting daily activities. The spatial information management mechanism efficiently partitions the space into regions using Geohash. The dynamic point cloud integration mechanism incorporates point clouds collected by 3D-MCS into UDTs through global and local point cloud registration. Finally, we evaluated the proposed framework through real-world experiments. We verified the effectiveness of the proposed 3D-MCS models from the perspectives of subjective evaluation and data collection and analysis. Furthermore, we analyzed the performance of the dynamic point cloud integration using a dataset.
\end{abstract}


\section{Introduction}

\IEEEPARstart{I}{n} smart cities, information and communication technologies are expected to address various urban challenges and enable the creation of new services \cite{bib:sc-survey}. Digital twin (DT) technology is crucial for implementing diverse urban services comprehensively, leading to sustainable urban digital twins (UDTs) \cite{bib:udt-01}.

Constructing and maintaining UDTs requires the continuous collection of large-scale spatial information as spatial DTs~\cite{bib:spatial-dt}. To achieve this, citizen-centric and participatory approaches have been proposed \cite{bib:d2ecosys}. Mobile crowdsensing (MCS)~\cite{bib:mcs-survey}, which collects data via mobile devices carried by city residents, is a promising approach. As modern mobile devices become more powerful, the types of data they can collect have grown more precise and diverse, enabling real-time data acquisition. In particular, the proliferation of mobile devices equipped with Light Detection and Ranging (LiDAR) sensors enhances spatial data collection, facilitating the construction and maintenance of UDTs.

Typically, MCS requires a large number of participants. However, challenges such as computational, communication, and power resource limitations, as well as privacy concerns, can reduce user motivation. Therefore, designing effective incentive mechanisms to encourage participation is essential \cite{bib:mcs-incentive-survey}. Additionally, when implementing MCS for spatial information to maintain UDTs, a comprehensive framework is necessary, encompassing not only the MCS mechanism itself but also the integration and management of collected spatial data.

In this article, we present a 3D mobile crowdsensing (3D-MCS) framework for spatial information collection and management, aiming to support sustainable UDTs in smart cities.

\section{3D-MCS Framework for Sustainable UDTs}

\begin{figure}[!b]
  \centering
  \includegraphics[width=0.90\columnwidth]{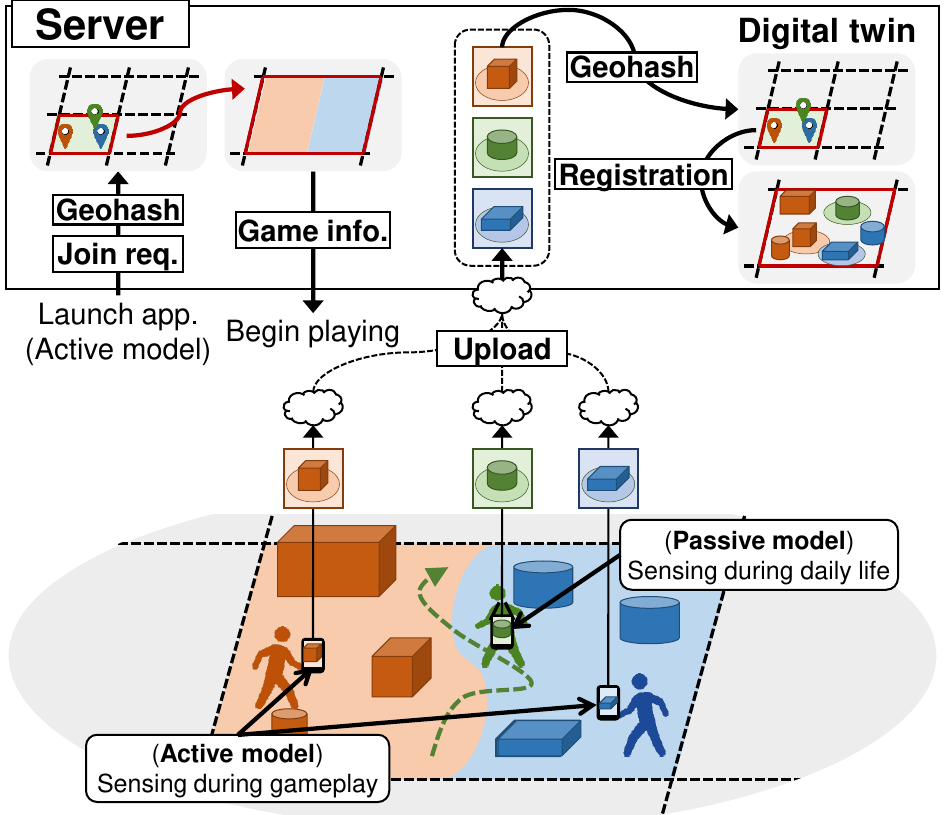}
  \caption{Overview of the 3D-MCS framework.}
  \label{fig:prop-concept-architecture}
\end{figure}

We propose a 3D-MCS framework for collecting spatial information in urban environments using LiDAR-equipped mobile devices carried by city residents or visitors, contributing to sustainable UDTs. This article mainly presents the comprehensive design of the 3D-MCS framework, building on extended ideas from preliminary versions of its components \cite{bib:3d-mcs-active,bib:3d-mcs-passive,bib:3d-mcs-registration,bib:3d-mcs-visualizer}.

\begin{figure}[!t]
  \centering
  \includegraphics[width=0.98\columnwidth]{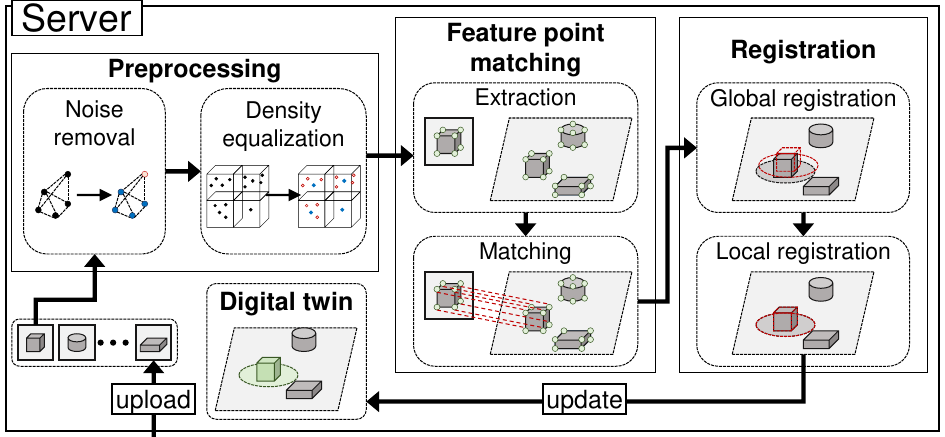}
  \caption{Overview of the procedure for integrating point clouds collected by the 3D-MCS framework.}
  \label{fig:prop-concept-registration}
\end{figure}

Figures~\ref{fig:prop-concept-architecture} and \ref{fig:prop-concept-registration} illustrate the overall architecture of the proposed 3D-MCS framework and the process of integrating point clouds collected through this framework. The framework comprises four key mechanisms: (1) the 3D-MCS mechanism based on active and passive sensing models, (2) the spatial information management mechanism leveraging Geohash, (3) the dynamic point cloud integration mechanism, and (4) the web-based real-time visualizer for 3D-MCS and UDTs.

In the 3D-MCS mechanism, The active model features a gamified spatial sensing approach using augmented reality (AR) technology to provide gentle nudges, encouraging participants to move to locations where data collection is needed. The passive model offers a wearable spatial sensing method that does not interfere with daily activities.

The spatial information management mechanism leverages Geohash, a geocoding system that encodes geographic coordinates, to systematically manage both the collected point cloud data and the game information used by the active gamified approach.

In the dynamic point cloud integration mechanism, the point cloud data collected by participants is dynamically merged into the city-scale point cloud maintained by the server, which serves as the UDT. This integration is performed by combining global and local point cloud registration techniques, enabling dynamic estimation of the appropriate alignment and integration positions within the UDT.

In the real-time visualizer, people can visually inspect the point clouds obtained via 3D-MCS and UDTs, while also accessing easily understandable supplementary information provided by various machine-learning-enabled tasks, such as predictions and detections, to encourage behavioral changes.

\subsection{Active 3D-MCS: Gamified Model}

\begin{figure}[!t]
  \centering
  \includegraphics[width=0.90\columnwidth]{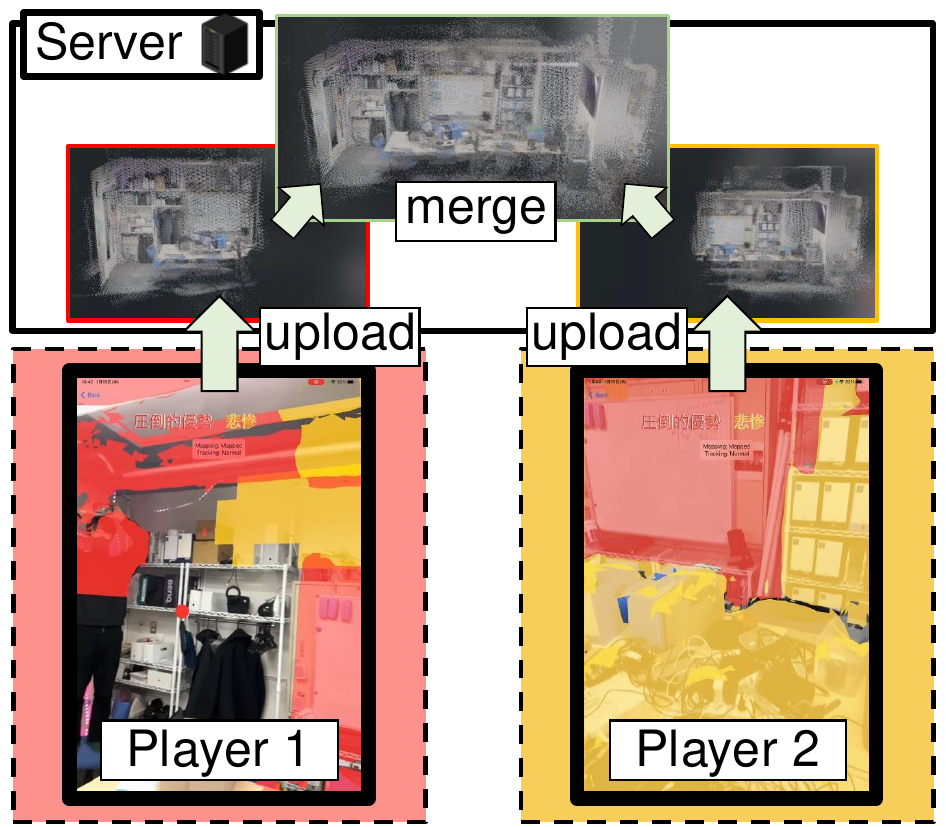}
  \caption{Operational example of the active gamified 3D-MCS model, showing actual game screens from the developed application alongside point clouds captured by the application.}
  \label{fig:prop-concept-active-example}
\end{figure}

In the active sensing model of the proposed 3D-MCS mechanism, we introduce an AR territory coloring game as a gamified element to incentivize participants to engage in 3D spatial sensing. This preliminary idea was proposed in \cite{bib:3d-mcs-active}. Figure~\ref{fig:prop-concept-active-example} shows an operational example of our gamified model. While participants play the coloring game in the AR space, their mobile devices continuously collect point cloud data via onboard LiDAR sensors.

First, a participant’s device sends a join request to the server, which then responds with the game information. After receiving the game information, the participant joins the game and becomes a player on one of several teams. Players can interact with the AR interface and participate in the AR territory coloring game.

\begin{figure}[!t]
  \centering
  \includegraphics[width=0.98\columnwidth]{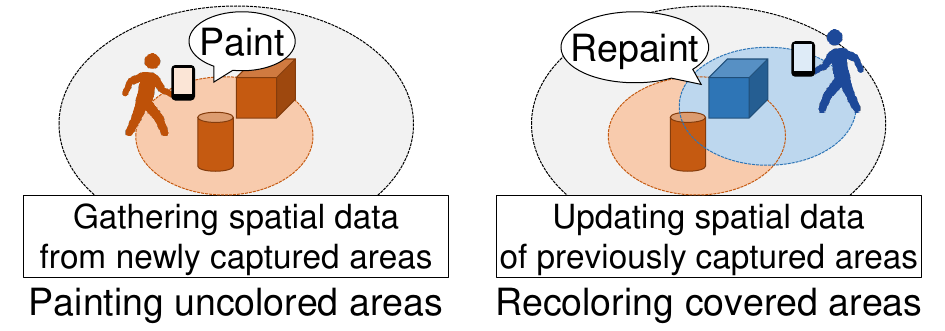}
  \caption{Key concept for motivating players to scan both uncaptured and previously captured areas using our gamification model.}
  \label{fig:prop-concept-active-model}
\end{figure}

\begin{figure}[!t]
  \centering
  \includegraphics[width=0.98\columnwidth]{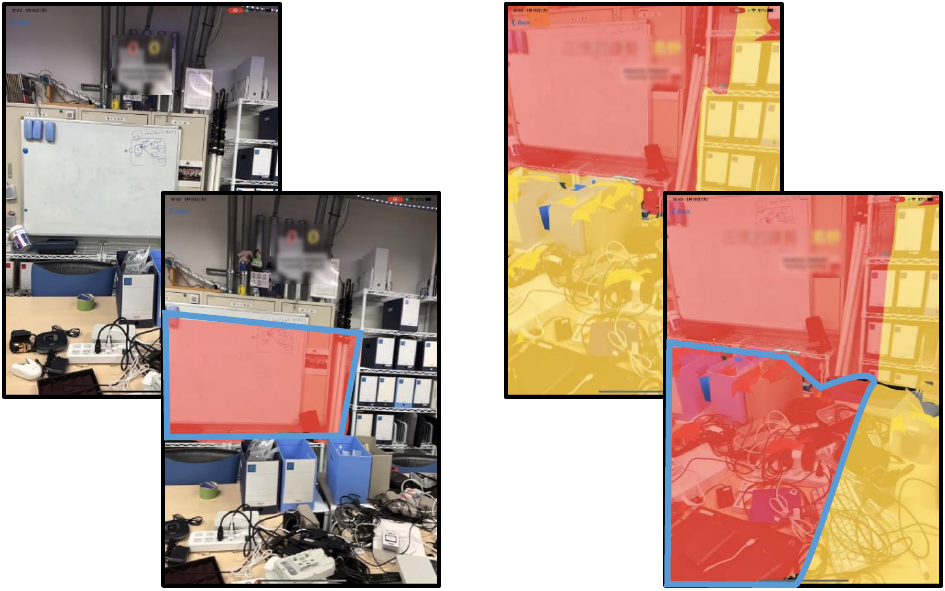}
  \caption{Screenshots of the developed application illustrating scenarios of painting uncolored areas and repainting previously colored areas in the AR space.}
  \label{fig:prop-concept-active-gui}
\end{figure}

Players can color physical objects and walls within the real-world space through the AR interface. The colored regions become territories belonging to their team, contributing to the team’s overall score. The territories acquired by each player are shared in real time among all participants and visualized within the AR space. This real-time sharing encourages players to explore and identify uncolored regions.

Figures~\ref{fig:prop-concept-active-model} and \ref{fig:prop-concept-active-gui} illustrate the key idea of motivating players to scan both uncaptured and previously captured areas. By painting uncolored regions, players collect point cloud data from uncaptured areas. Additionally, players can recolor areas that were previously colored by other teams, thereby collecting updated point cloud data and refreshing spatial information.

\subsection{Passive 3D-MCS: Wearable Model}

When people capture spatial information without relying on specific spots or objects, it is desirable to acquire point cloud data without interfering with participants' behavior. Therefore, we propose a passive wearable 3D-MCS model in which people wear LiDAR-enabled mobile devices around their necks or place them in a breast pocket as a wearable LiDAR sensor. The preliminary idea was proposed in \cite{bib:3d-mcs-passive}.

First, once people participate in the system, they wear the LiDAR-enabled mobile device and regularly capture point cloud data using their devices without consciously focusing on the sensing behavior. The mobile device periodically uploads the captured point cloud data and other sensing information to the server. This model can thus acquire spatial information from within the participants' daily life sphere in urban spaces, such as the roads they typically walk on.

\subsection{Geohash-based Spatial Information Management}

In the proposed framework, the volume of collected point clouds and/or AR game information scales with the size of the urban area. To efficiently manage this data, the proposed framework adopts a spatial information management mechanism based on Geohash, which encodes geographic coordinates into short strings. 

When newly participating in the active gamified model, the server sends AR game information to participants corresponding to the Geohash region derived from their current location. In addition, the server continuously manages and collects point clouds sent from participants of both active and passive 3D-MCS models based on the Geohash code corresponding to their location, either during gameplay or while carrying out daily activities. As a result, the spatial information is independent of other Geohash regions, making the mechanism scalable, as it can manage each small-scale region with multiple servers in a distributed manner. Additionally, it enhances the accuracy and speed of point cloud registration in the dynamic integration mechanism into the UDT.

\subsection{Dynamic Point Cloud Integration into UDTs}

To maintain the UDT, point clouds collected through 3D-MCS must be dynamically integrated into the appropriate locations within the large-scale point cloud representing the UDT. We propose a dynamic point cloud integration mechanism based on global and local registration algorithms. Details of our algorithm have been introduced in \cite{bib:3d-mcs-registration}.

When the server receives a point cloud from a participant, it first removes noise using a statistical outlier removal filter, and then equalizes the point density using a voxel grid filter as a preprocessing step. Subsequently, the server extracts feature points and establishes a correspondence set between the collected point cloud and the partial point cloud included in the UDT during the feature point matching step. Next, the server applies random sample consensus (RANSAC) to globally estimate the appropriate position and orientation of the collected point cloud. If an initial alignment is found, local registration is performed using iterative closest point (ICP) to refine the alignment by minimizing the distance between corresponding points. Through this process, the collected point cloud is dynamically and accurately integrated into the UDT.

\subsection{3D-MCS and UDT Visualizer}

\begin{figure}[!t]
  \centering
  \includegraphics[width=0.95\columnwidth]{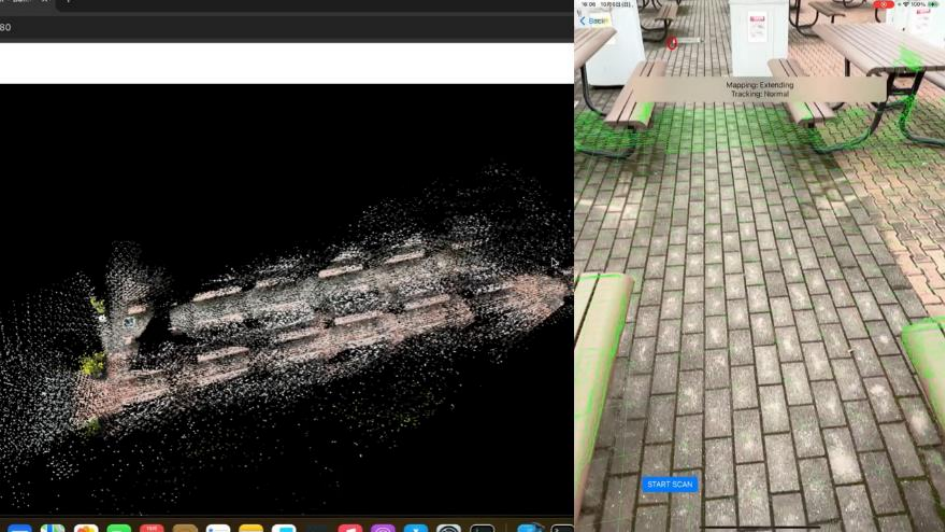}
  \caption{Screenshots of the visualizer (left) and game screen (right) captured in a real-world environment. The visualizer displays the collected point clouds and UDTs on the server in real time.}
\label{fig:prop-visualizer-example}
\end{figure}

Visualizing the point clouds obtained via 3D-MCS and UDT is important for examining their status and providing understandable supplementary information within UDTs. This additional information can include results from various machine-learning-enabled tasks, such as predictions and detections, aimed at encouraging behavioral changes in people. From another perspective, human-in-the-loop approaches can be implemented for the alignment and integration processes of point clouds using the visualized results.

To address these challenges, we propose a web-based real-time visualizer for the 3D-MCS framework. A preliminary version of this concept was introduced in \cite{bib:3d-mcs-visualizer}. Figure~\ref{fig:prop-visualizer-example} shows actual screenshots of our visualizer alongside the corresponding game screen. This mechanism supports real-time visualization across various use cases, including: (1) the incremental approach, (2) the frame-based approach, and (3) the situational approach. The incremental approach continuously visualizes the collected point clouds while retaining previously displayed data. The frame-based approach displays the current point cloud frame in real time, similar to visualization applications designed for individual LiDAR devices. The situational approach colors and visualizes point clouds based on service-specific conditions.

\subsection{Implementation}

We developed the proposed framework consisting of client applications and server software. The client applications for the active and passive 3D-MCS models were implemented as iOS apps targeting LiDAR-enabled iPhone and iPad models, using ARKit and MetalKit. These applications periodically upload partial point clouds to the server once the stored points exceed 200,000. The server software was developed using Node.js and the Express framework to manage captured point cloud data and AR game information, organized by Geohash regions, via an HTTP REST API. The proposed visualizer was implemented as a web-based application using Three.js and therefore it is available on web browsers It communicates with the server using the REST API to obtain point clouds and the UDT and visualizes them. Point cloud data is stored in Polygon File Format (PLY), which allows flexible modification of per-point information. In our implementation, client applications can save PLY files in either text or binary format. Each PLY file includes a header followed by data for each point, such as 3D coordinates, RGBA color, confidence, depth, roll, angular velocity, self-localized position, and other attributes.

\section{Real-World Experiments and Performance Analysis}
\label{sec:experiment}

\begin{figure}[!t]
  \centering
  \begin{minipage}{0.55\columnwidth}
    \centering
    \includegraphics[width=0.98\columnwidth]{fig/exp-devices.pdf}
    {\footnotesize(a)~Experimental equipment}
  \end{minipage}
  \begin{minipage}{0.36\columnwidth}
    \centering
    \includegraphics[width=0.98\columnwidth]{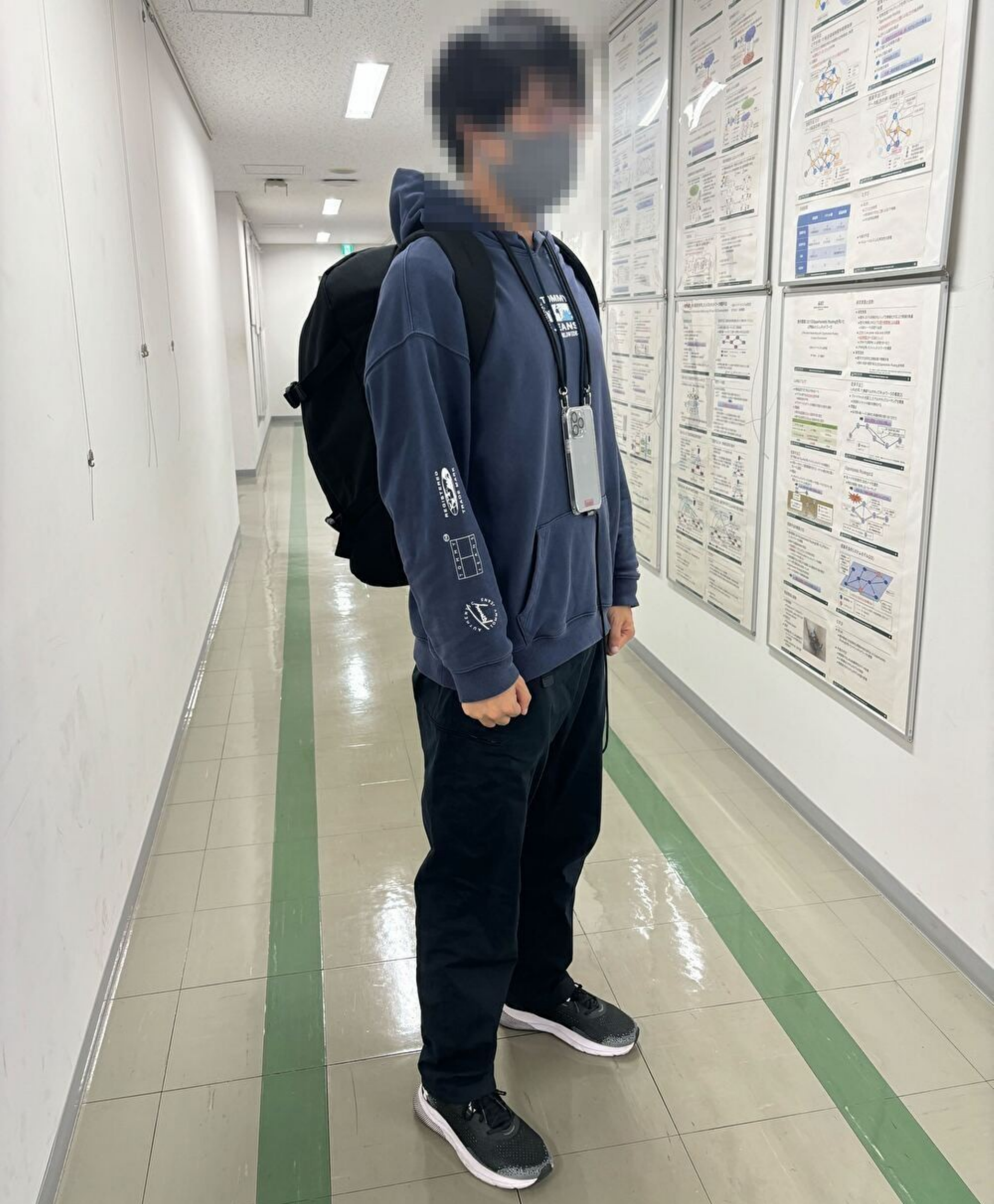}
    {\footnotesize(b)~Experimental setup}
  \end{minipage}
  \caption{Experimental equipment and setup: (a) A backpack containing a server, wireless LAN router, and mobile battery. (b) A subject carrying the backpack and either wearing a LiDAR-enabled smartphone around their neck or holding it in their hand.}
  \label{fig:exp-subject-setup}
\end{figure}

We comprehensively analyzed performances of our 3D-MCS framework through real-world experiments. Experiments were conducted at the Omiya campus, Shibaura Institute of Technology, Japan. Experiments adopted Apple iPhone 15 Pro and Apple iPad Pro as LiDAR-enabled mobile devices to run the 3D-MCS applications. The management server of our 3D-MCS framework runs on a MacBook Air in outdoor experiments. The mobile device and server were connected to a wireless local area network via the GL.iNet GL-XE300 mobile router. Figure~\ref{fig:exp-subject-setup} shows that the subject carried a backpack containing the server and the mobile router, while either wearing a LiDAR-enabled smartphone around their neck or holding it in their hand, depending on active or passive 3D-MCS models. The mobile device and server were connected via a USB cable for analyzing logs only.

\subsection{Point Cloud Collection via Active Gamified Model}

\begin{figure}[!t]
  \centering
  \includegraphics[width=0.80\columnwidth]{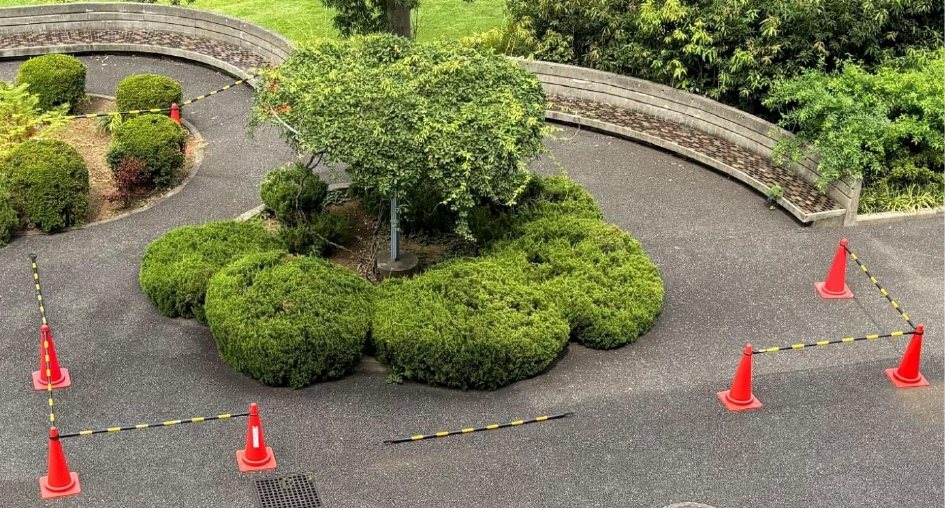}
  \caption{Experiment locaiton.}
  \label{fig:exp-active-location}
\end{figure}

To verify the quality and status of captured point clouds via the active gamified model, we conducted experiments under real-world conditions on the Omiya campus, Shibaura Institute of Technology, Japan. We conducted a basic functionality test using the proposed system implementation in the open area beside Building 3 on the Omiya campus of our university. Figure~\ref{fig:exp-active-location} shows the experiment location. In this experiment, the subject scanned an 8\,m $\times$ 10\,m area enclosed by traffic cones over a period of approximately 200\,s. We analyzed the collected point cloud and the AR game information during this period.

\begin{figure}[!t]
  \centering
  \includegraphics[width=0.80\columnwidth]{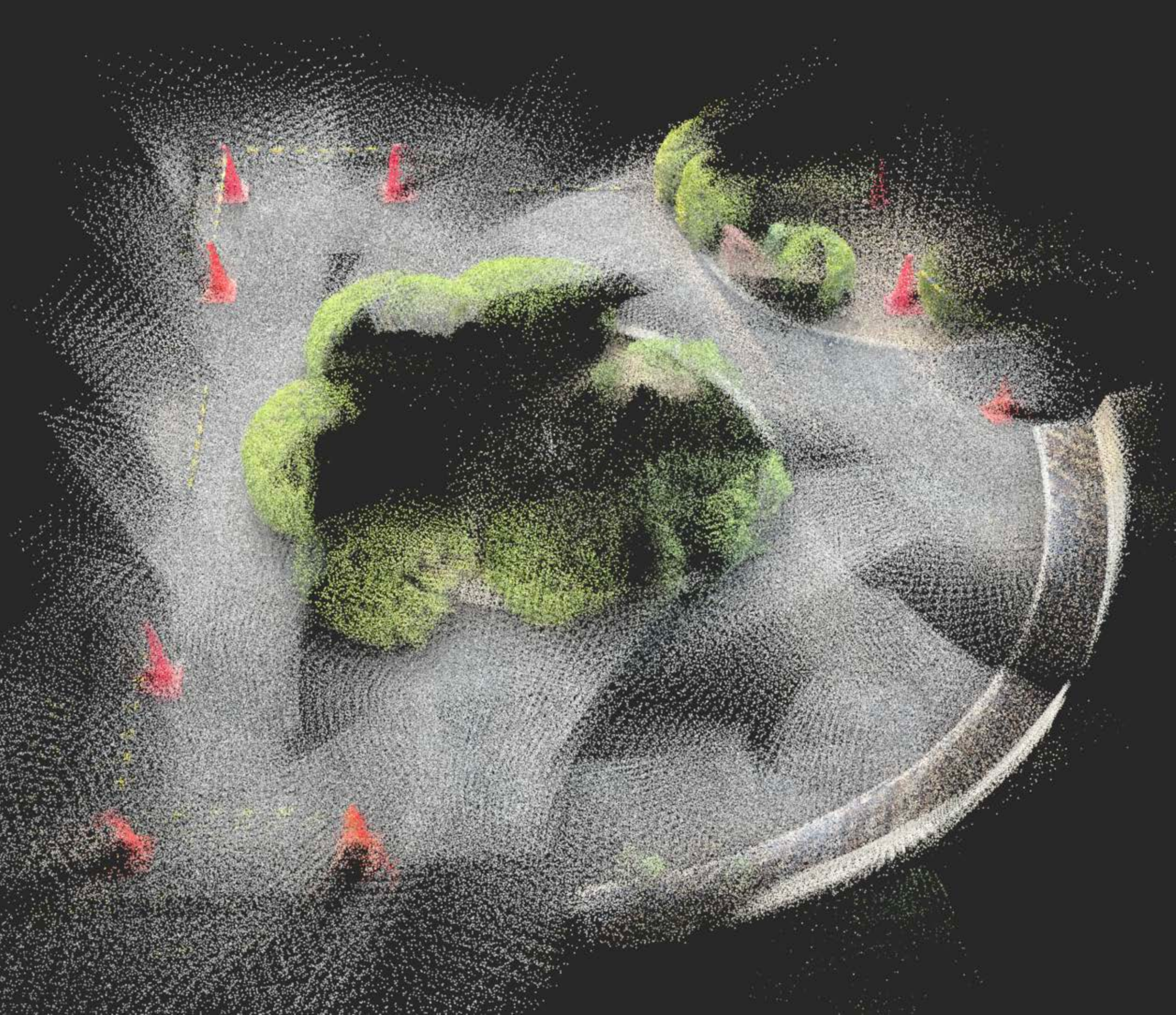}
  \caption{Visualized example of collected point cloud by single sensing attempt.}
  \label{fig:exp-active-pc}
\end{figure}

\begin{figure}[!t]
  \centering
  \includegraphics[width=0.80\columnwidth]{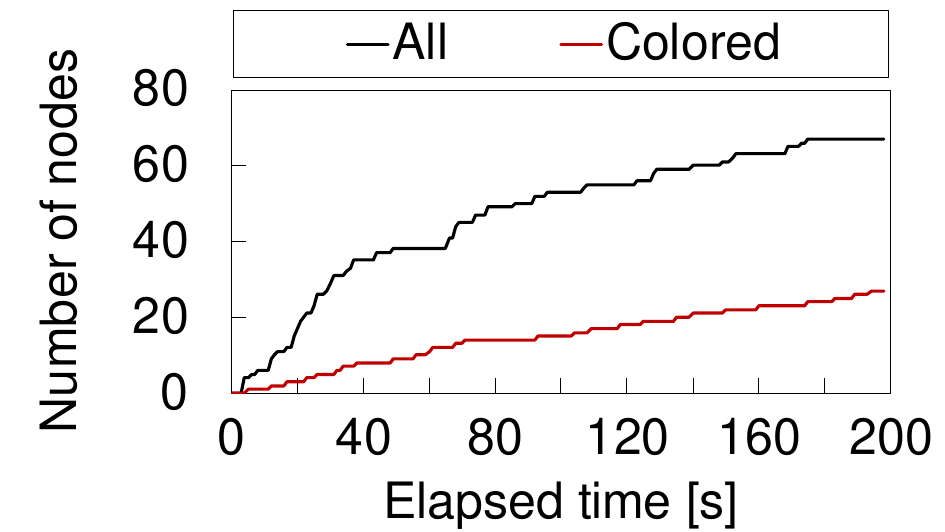}
  \caption{Transition of numbers of uncolored/colored AR nodes.}
  \label{fig:exp-active-arnode}
\end{figure}

Figure~\ref{fig:exp-active-pc} visualizes the point cloud data collected by the active gamified model, demonstrating that the data was appropriately captured. Figure~\ref{fig:exp-active-arnode} shows the time transition in the number of AR nodes, which are components of the AR map. As time progresses, both the number of AR nodes and the number of colored AR nodes increase, indicating that the real-world space is being correctly recognized in AR space and that AR nodes are being successfully painted.

\subsection{Subjective Evaluation for Active Gamified model}

To evaluate the effectiveness of the active gamified model, we conducted a subjective evaluation from the perspective of user experience. In this evaluation, we compared two applications: one based on an AR territory coloring game and the other that simply visualizes point clouds oriented towards the camera without any game elements.

The experiment was conducted in our laboratory with 30 subjects, all of whom were students. The experimenter explained how to use the applications and obtained verbal informed consent before each subject used the applications for one minute. Subsequently, they answered a questionnaire, which is composed of six questions to assess the motivational impact of the gamification on user behavior during sensing. The first four questions Q1-Q4 were designed for verifying impact of gamification on motivation with reference to the game experience questionnaire (GEQ) \cite{bib:geq}. The last two questions Q5 and Q6 were designed for verifying impact on behavioral changes by the territory coloring game.

Q1--Q4 are based on the core module of GEQ: positive affect (Q1 and Q2) and negative affect (Q3 and Q4) as follows: (Q1) ``The subject felt the experience was enjoyable''; (Q2) ``The subject felt satisfaction or a sense of accomplishment''; (Q3) ``The subject felt the experience was troublesome''; and (Q4) ``The subject felt the experience was boring.'' Q5--Q6 focus on the impact on behavioral change as follows: (Q5) ``Did you consciously try to color areas that had not been colored?''; and (Q6) ``Did you consciously try to recolor areas that had been colored by the opposing team?'' Note that the questions have been translated from Japanese.

\begin{table}[!t]
  \begin{center}
  \caption{Subjective evaluation results of Q1--Q6. Q5 and Q6 are designed specifically to assess gamification, and therefore there are no results for the condition without gamification.}
  \label{tab:exp-active-subjective}
  \begin{tabular}{l|cc}
    \hline
     & \multicolumn{2}{c}{Average score with STD}\\
     & w/o Gamification & w/ Gamification \\
    \hline
     Q1 ($\uparrow$) & $3.17 \pm 1.32$ & $4.47 \pm 0.88$  \\
     Q2 ($\uparrow$) & $3.17 \pm 1.37$ & $4.07 \pm 0.93$   \\
     Q3 ($\downarrow$) & $2.30 \pm 1.16$ & $1.83 \pm 1.21$  \\
     Q4 ($\downarrow$) & $2.47 \pm 1.20$ & $1.53 \pm 0.99$ \\
     Q5 ($\uparrow$) & N/A & $4.03 \pm 1.05$ \\
     Q6 ($\uparrow$) & N/A & $3.93 \pm 1.06$ \\
   \hline
  \end{tabular}
  \end{center}
\end{table}

Table~\ref{tab:exp-active-subjective} shows the evaluation results of Q1--Q6. The questionnaire employed a five-point Likert scale, where higher scores indicated greater agreement or a stronger feeling. Q1 and Q2 indicate that subjects felt more positive affect, while Q3 and Q4 suggest they experienced less negative affect when compared to the condition without gamification. This result demonstrates that the introduction of gamification led to a more positive perception of mobile sensing, indicating that gamification was effective in enhancing motivation. In Q5 and Q6, subjects gave high ratings, suggesting that the coloring-based territory game likely encouraged behavioral changes in participants within our active gamified model.

\begin{figure}[!b]
  \centering
  \includegraphics[width=0.95\linewidth]{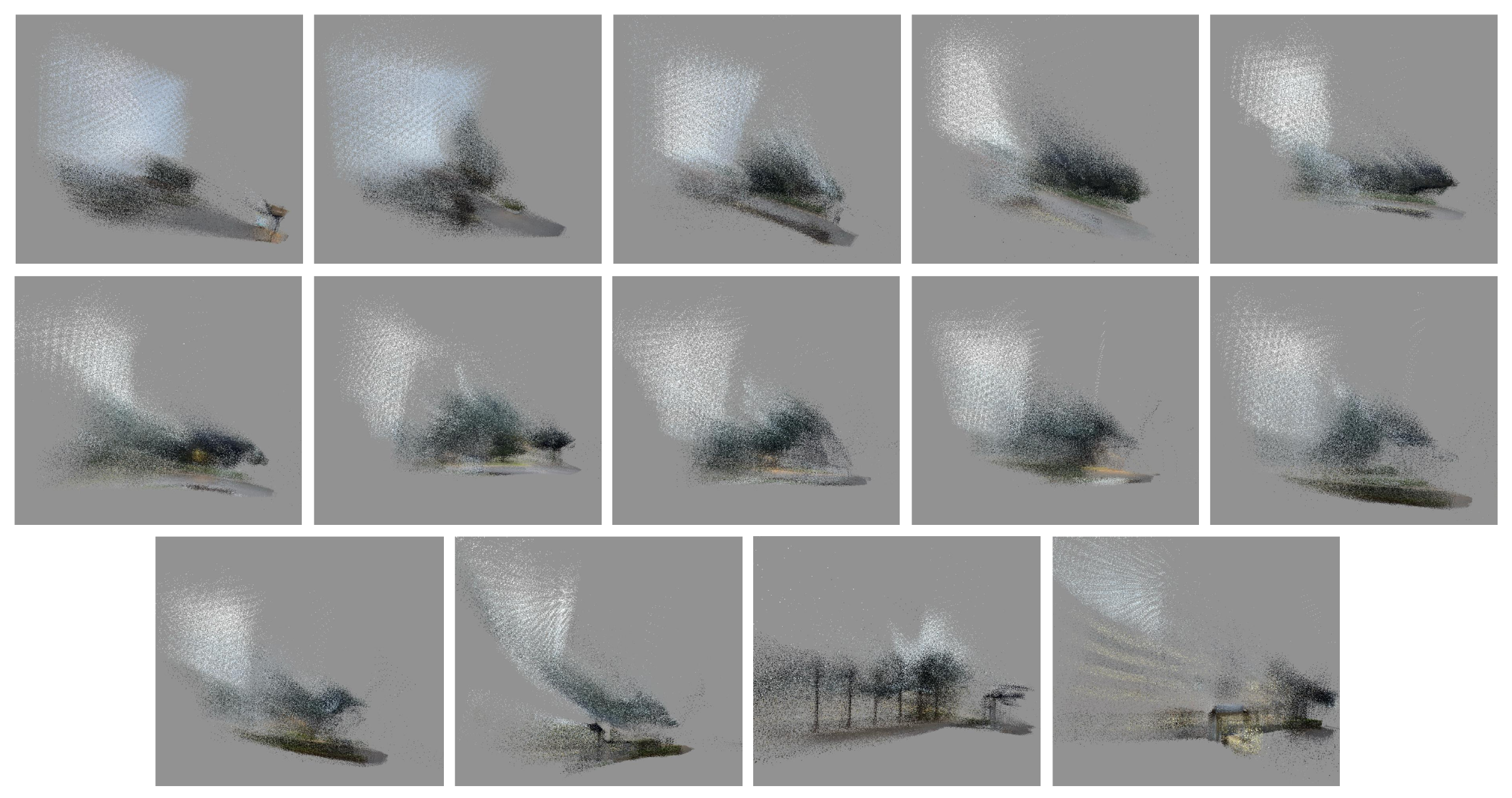}
  \\ {\vspace{-0.0em}\footnotesize(a)~Sequentially collected point clouds\vspace{1.0em}}
  \includegraphics[width=0.95\linewidth]{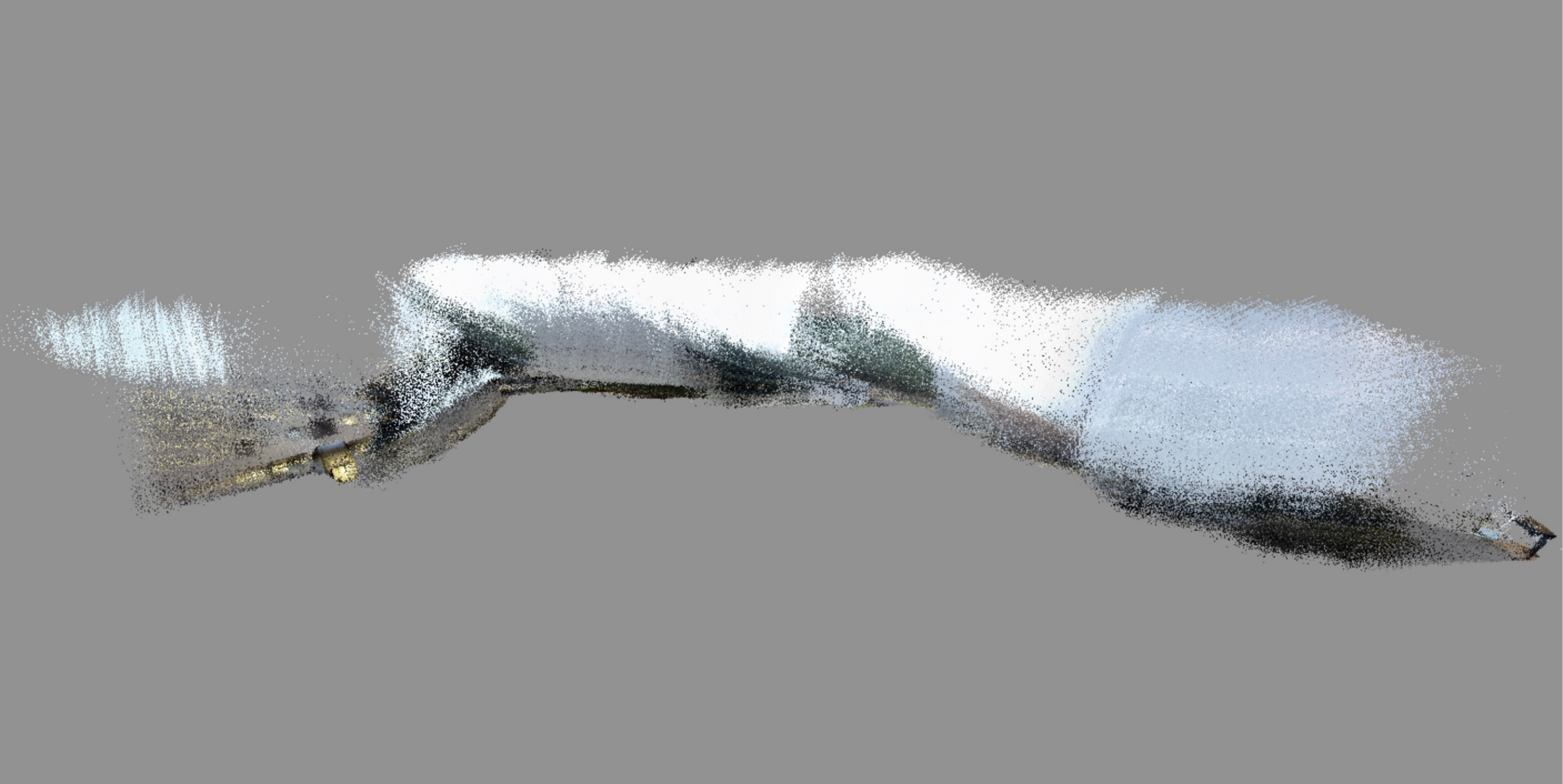}
  \\ {\vspace{-0.0em}\footnotesize(b)~Merged point cloud}
  \caption{Point clouds collected while walking from the campus store to the 5th building: (a) All point clouds are stored separately and sent to the server in a time-series manner. (b) Sequentially collected point clouds are aligned in the same coordinate system using self-localization and mapping provided by the ARKit framework.}
  \label{fig:exp-passive-walking-pcs}
\end{figure}

\subsection{Point Cloud Collection via Passive Wearable Model}

To verify the feasibility of the proposed system, we conducted an experiment to collect point cloud via the passive wearable model in our university campus. In this experiment, a subject walked along a path from the campus store to the 5th Building in our campus, while collecting point cloud data using a mobile device. 

Figure~\ref{fig:exp-passive-walking-pcs} visualizes the sequential 14 point cloud samples and merged one using our web-based visualizer~\cite{bib:3d-mcs-visualizer} collected during the experiment. The data size of each point cloud sample was approximately 30\,MB and the size of merged one was 438.2\,MB in text-based format. The visualized results indicate that the passive wearable model successfully captured the real-world space, however, some unreliable points are included in the point cloud. 

\subsection{Impact of Filtering Process of Unreliable Points and Properties}

\begin{figure}[!t]
  \centering
  \includegraphics[width=0.70\columnwidth]{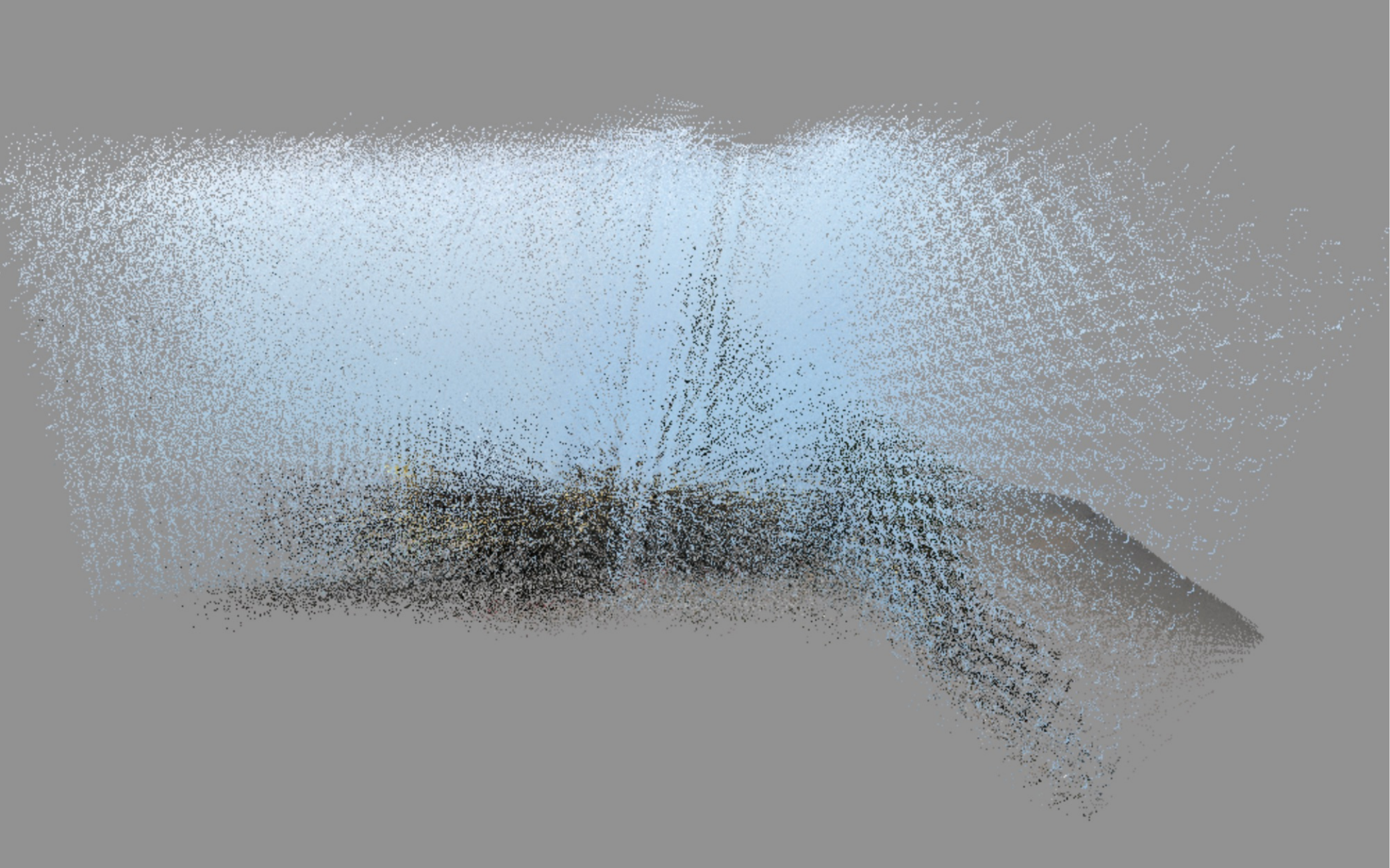}
  \\ {\footnotesize(a)~Raw point cloud without any filtering applied\vspace{1.0em}}
  \includegraphics[width=0.70\columnwidth]{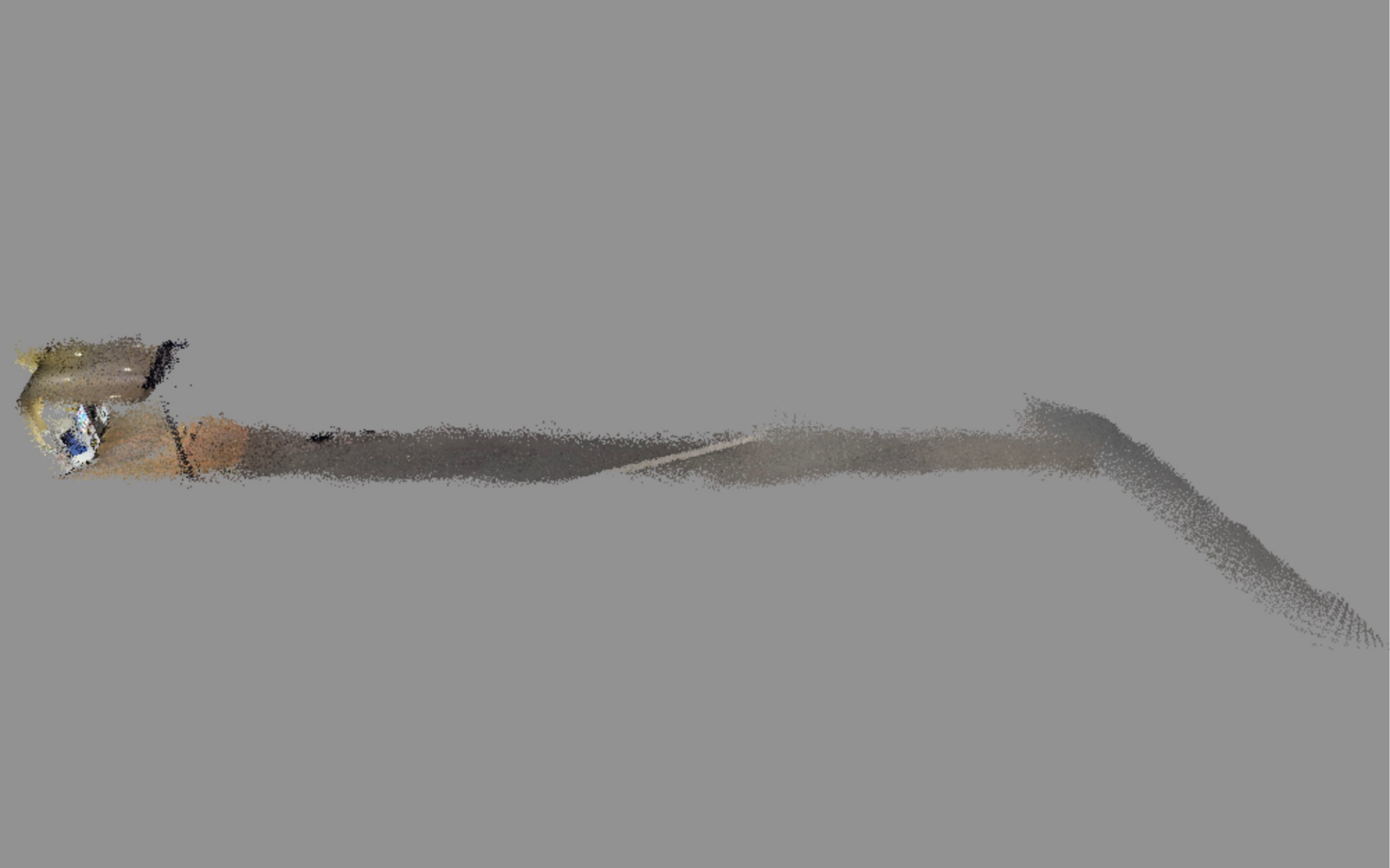}
  \\ {\footnotesize(b)~Filtered result by Confidence and depth}
  \caption{Visualized point clouds with filtering process of unreliable points.}
  \label{fig:exp-passive-filtering-visualization}
\end{figure}

Filtering unreliable points from the collected point cloud is important to improve the success rate of point cloud registration and keep the UDT accurate. We conducted an experiment using point properties to filter out unreliable points and examined the differences in the appearance between the unfiltered and filtered ones.  We collected point cloud data by walking from the main gate to the campus store on our campus. To remove unreliable points from the collected data, we used parameters such as the confidence level of the collected points and the depth from the LiDAR sensor to each point. The confidence level is defined as a three level scale from zero to two, where a higher level is more confident. In this experiment, we removed points that were over 5\,m away from the LiDAR sensor and those with a confidence level of zero, which is the lowest confidence level.

\begin{table}[!t]
  \caption{Number of points and data sizes of the point cloud in text and binary formats with/without filtering by confidence level and depth.}
  \centering
  \begin{tabular}{cc|rrr}
    \hline
    Confidence & Depth & Num.\ pts & Text & Binary \\ \hline
     & & 1,033,942 & 165.4\,MB & 59.2\,MB\\ 
    \checkmark & & 708,784 & 113.4\,MB & 40.6\,MB\\
     & \checkmark & 303,283 & 48.6\,MB & 17.4\,MB\\ 
    \checkmark & \checkmark & 202,334 & 32.5\,MB & 11.6\,MB\\ \hline
  \end{tabular}
  \label{tab:exp-passive-filtering-datasize}
\end{table}

\begin{table}[!t]
  \caption{Data size in binary format occupied by each property within the point cloud.}
  \centering
  \begin{tabular}{r|r}
    \hline
     & Data size \\ \hline
    3D coordinates & 11.83\,MB \\ 
    RGBA & 3.94\,MB \\
    Confidence & 3.94\,MB \\ 
    Depth & 3.94\,MB \\ 
    Others & 35.50\,MB \\
    Total & 59.16\,MB \\ \hline
  \end{tabular}
  \label{tab:exp-passive-breakdown}
\end{table}

Fig.~\ref{fig:exp-passive-filtering-visualization} visualizes point clouds with/without filtering processes. The filtering process can effectively remove sporadic unreliable points based on the confidence level and depth since the operational range of the iPhone LiDAR is only for a few meters. Therefore, points captured beyond this range tend to have lower accuracy. Next, Table~\ref{tab:exp-passive-filtering-datasize} compares the data size between the raw point cloud and filtered ones. We can confirm that filtering based on depth and confidence level effectively reduces the data size while removing unreliable points.

In the point cloud in our 3D-MCS framework, various properties are assigned to each point. Currently, the point cloud data container contains the 3D coordinates, RGBA, confidence, depth, as well as others (e.g., localized position, IMU, and etc.). Table~\ref{tab:exp-passive-breakdown} indicates that the data size is only 11.83\,MB when including 3D coordinates only. However, adding other properties significantly increases the sensing data size, and hence we need to appropriately adopt properties to effectively reduce the data size.

\subsection{Impact of Area Size on AR Map Information}

\begin{figure}[!t]
  \centering
  \includegraphics[width=0.90\columnwidth]{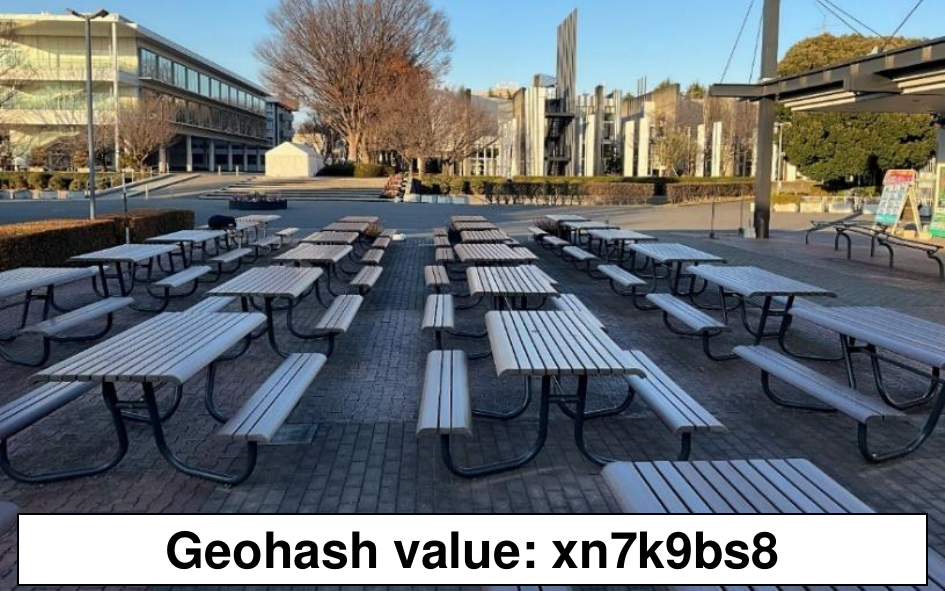}
  \caption{Experiment location designated by the Geohash precision level 8.}
  \label{fig:exp-geohash-location}
\end{figure}

We investigated the impact of the target area size on AR game and coloring information in an open plaza in front of a campus store at Omiya campus in Shibaura Institute of Technology. Figure~\ref{fig:exp-geohash-location} shows the experiment field, where specified by Geohash level 8 (19.03\,m$\times$30.90\,m). We compared two scenarios with full size and half size areas designated by the level 8. The subject repeatedly performed sensing using the proposed application with active sensing model. We then examined the volume and composition of the AR game data stored on the server.

\begin{figure}[!t]
  \centering
  \includegraphics[width=0.95\columnwidth]{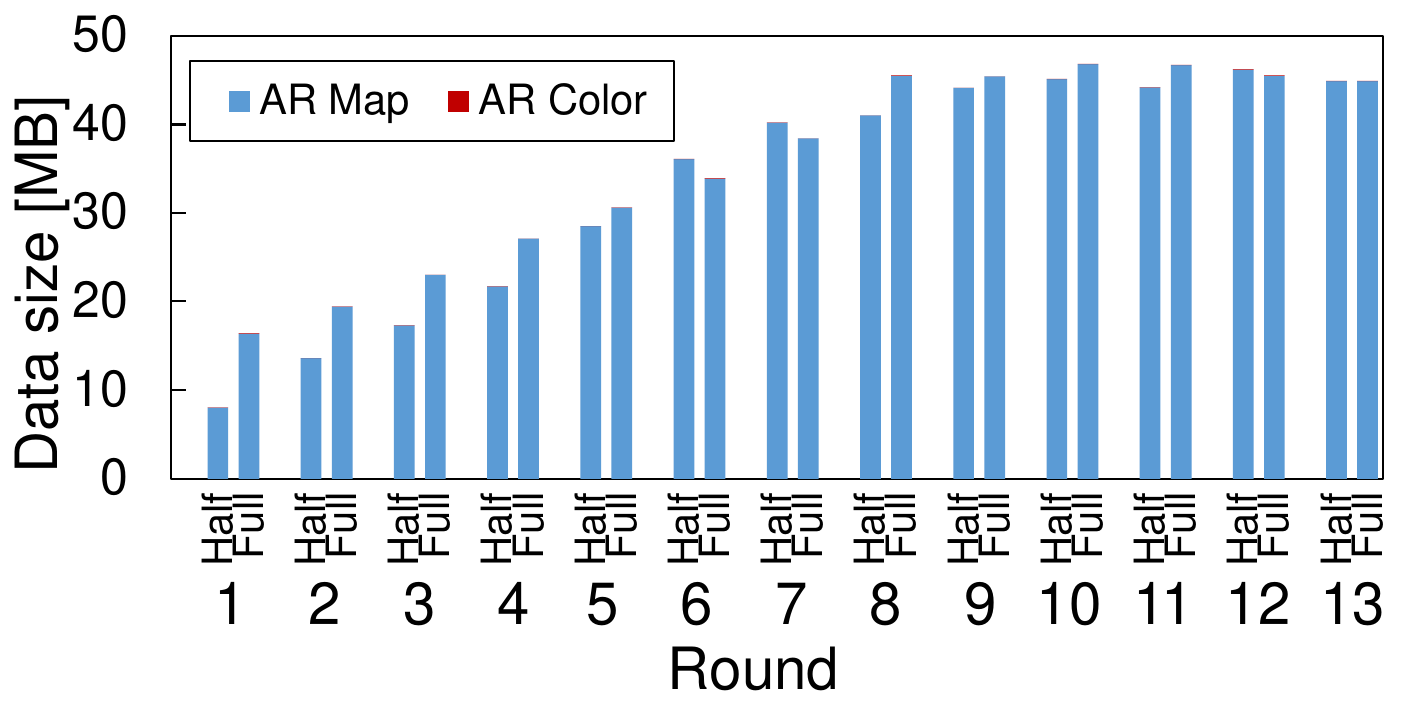}
  \caption{Data size of AR map and color information. We compared half and full size of Geohash region designated by Geohash precision level is 8.}
  \label{fig:exp-geohash-datasize}
\end{figure}

Figure~\ref{fig:exp-geohash-datasize} indicates the amount of AR game and coloring information. The data size is increased with each round. The data size eventually converged to a fixed value regardless of the area size. This suggests that ARKit may impose an upper limit on the size of the data saved as AR information. Thus, while it is possible to maintain a consistent data size irrespective of area size, this may affect the fidelity of the AR space reconstruction. 

\subsection{Accuracy and Processing Time of Dynamic Point Cloud Integration}

The accuracy and processing time of dynamic point cloud integration mechanism are evaluated using the VIRTUAL SHIZUOKA dataset \cite{bib:virtual-shizuoka}, a publicly available urban point cloud dataset provided by Shizuoka Prefecture. In this experiment, we assume that patial point clouds collected by 3D-MCS and they are merged with a global point cloud as a UDT to update those incomplete areas. We implemented the proposed mechanism using Python with the Open3D library. Note that we do not use preprocessing step in this experiment since we do not add additional noises in the dataset.

\begin{figure}[!t]
  \centering
  \includegraphics[width=0.95\columnwidth]{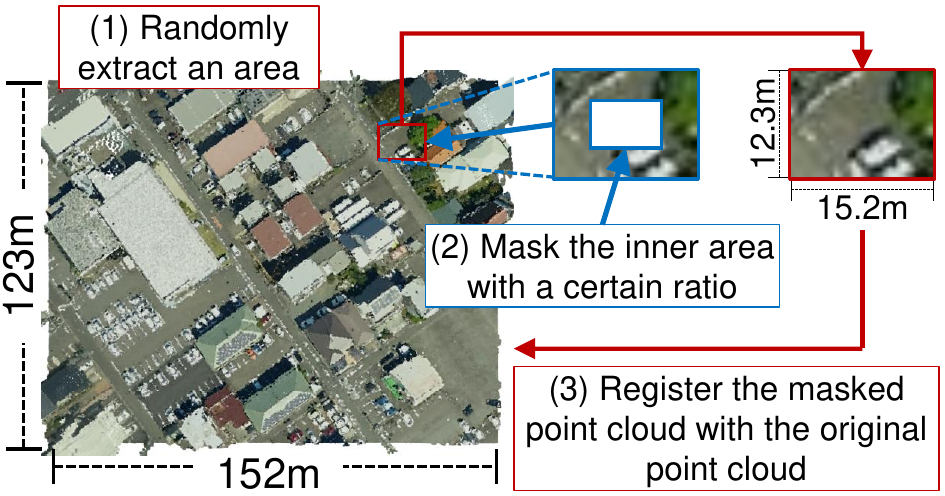}
  \caption{Experimental procedure of point cloud integraiton.}
  \label{fig:exp-integration-procedure}
\end{figure}

Fig.~\ref{fig:exp-integration-procedure} shows the dataset and the experimental procedure. First, a 123\,m $\times$ 152\,m region was extracted from the dataset as a section of UDT. Next, a 12.3\,m $\times$ 15.2\,m subregion was randomly selected from this region to assume the point cloud collected by the 3D-MCS. Subsequently, the extracted subregion was masked by a rectangular area centered at its midpoint. The size of the rectangular area was determined as a proportion of the side length of the subregion. Finally, we evaluated the registration success rate, root mean square error (RMSE), and processing time when the proposed system registered the extracted subregion within the region. We set the radius of the spherical region for FPFH feature calculation to five times the point spacing, the maximum number of nearest neighbor feature points selected for FPFH feature calculation to 100, and the threshold ratio for FPFH feature calculation to 0.9.

\begin{figure}[!t]
  \centering
  \includegraphics[width=0.75\columnwidth]{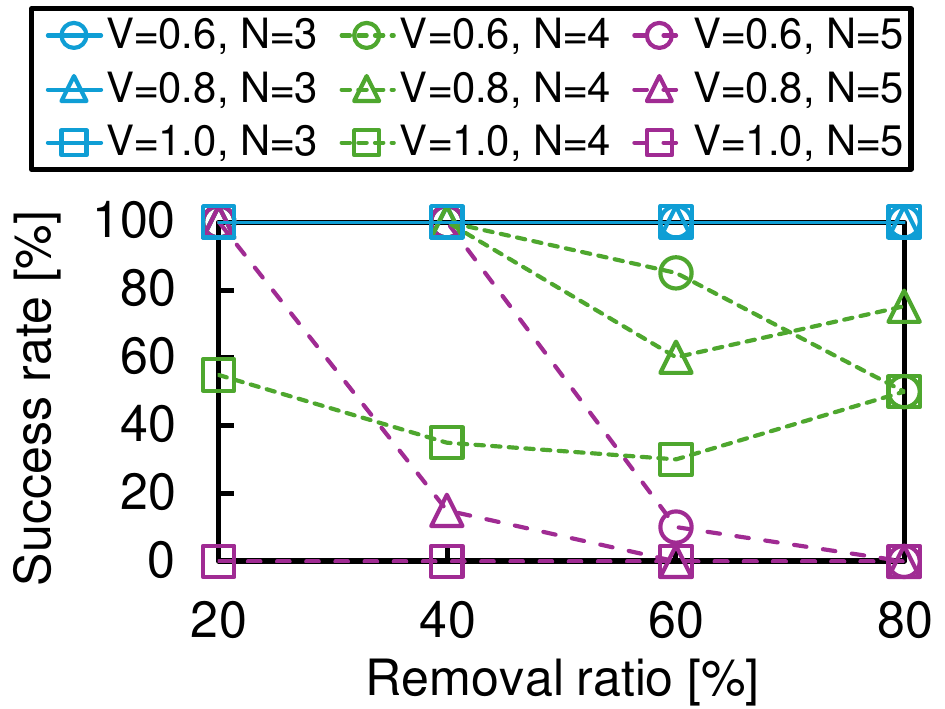}
  \caption{Registration success rate.}
  \label{fig:exp-integration-successrate}
\end{figure}

\begin{figure}[!t]
  \centering
  \includegraphics[width=0.80\columnwidth]{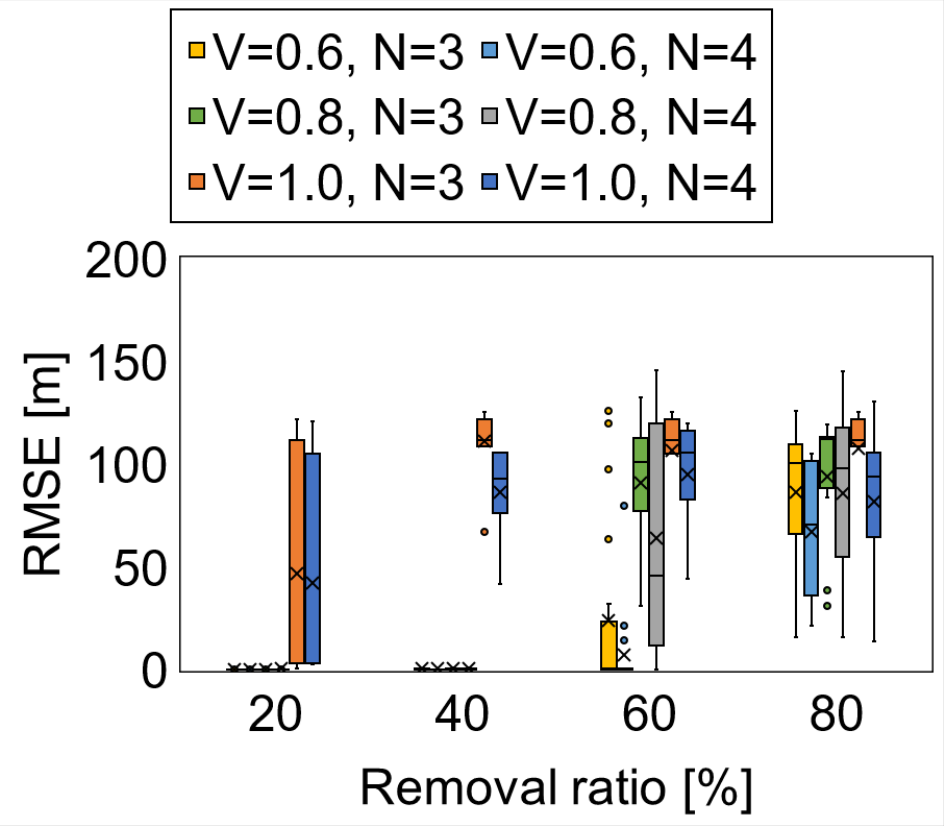}
  \caption{RMSE.}
  \label{fig:exp-integration-rmse}
\end{figure}

\begin{figure}[!t]
  \centering
  \includegraphics[width=0.80\columnwidth]{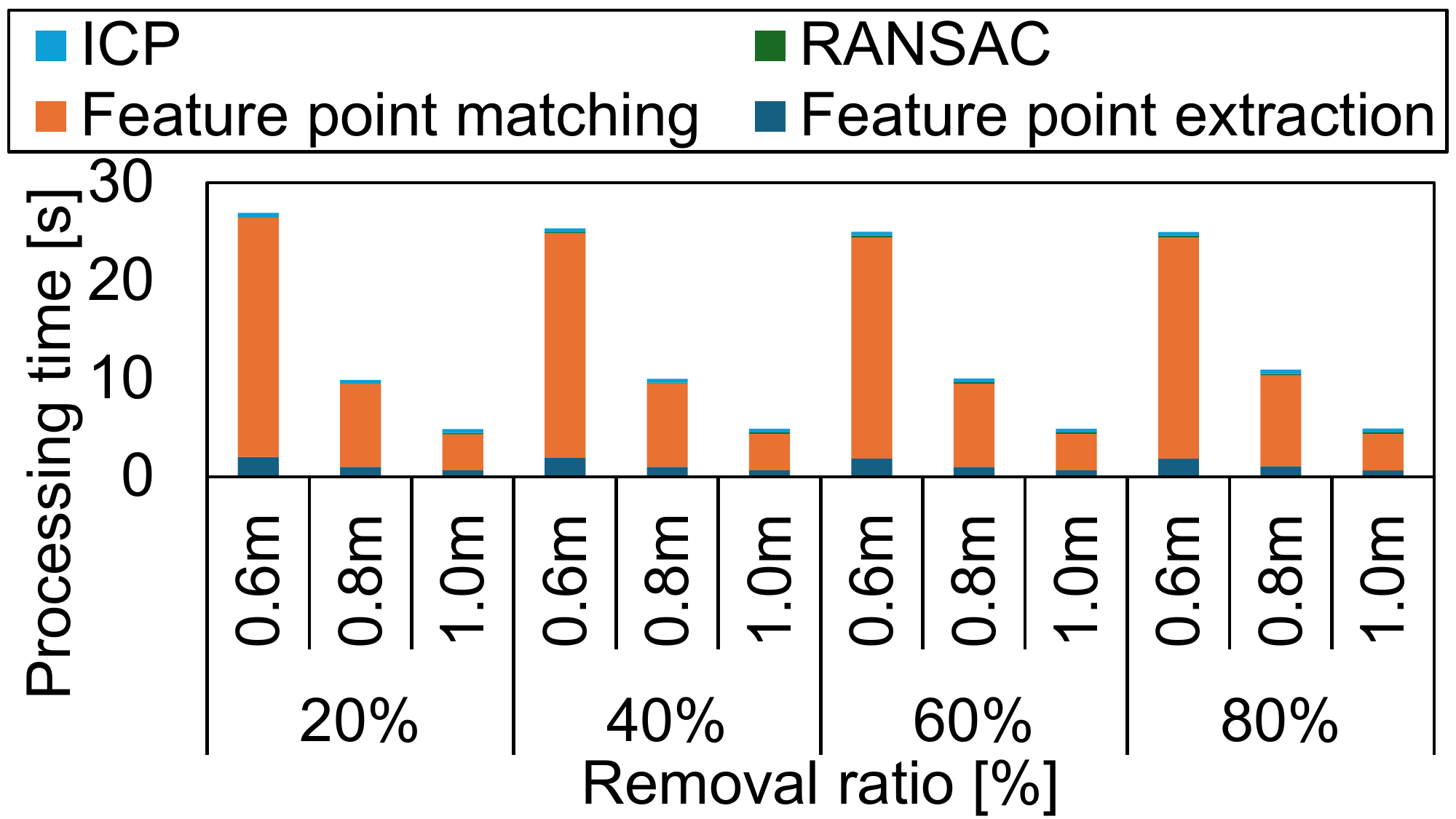}
  \caption{Processing time and its breakdown measured on Apple MacBook Pro (Intel Core i5-1038NG7, RAM 16\,GB, without GPU acceleration).}
  \label{fig:exp-integration-processing-time}
\end{figure}

Figure~\ref{fig:exp-integration-successrate} shows that the proposed mechanism ($N=3$) achieved a high success rate, but the success rate significantly decreased as $N$ increased, especially at high removal ratios, since RANSAC becomes less effective in discovering the transformation matrix. Furthermore, the success rate also decreased as $V$ increased. Figure~\ref{fig:exp-integration-rmse} shows the RMSE, which was calculated based on successful cases in RANSAC. The proposed mechanism achieved sufficient accuracy in point cloud registration when the removal ratios were 20\% and 40\% with $V=0.6$\,m and $V=0.8$\,m. Figure~\ref{fig:exp-integration-processing-time} shows the overall processing time and its breakdown. The feature matching process consumed a large proportion of the processing time, while the other processes had minimal impact. Additionally, the processing time depended on $V$, but was not affected by $N$.  In conclusion, the proposed mechanism can successfully register point clouds with low RMSE and short processing time when the removal ratio is approximately lower than 40\%.

\section{Conclusion}

In this article, we proposed a 3D-MCS framework aimed at realizing sustainable UDTs in smart cities. Additionally, we comprehensively evaluated the proposed framework through real-world experiments. In these experiments, we verified the effectiveness of the proposed 3D-MCS models from the perspectives of subjective evaluation and point cloud data collection and analysis. Furthermore, we analyzed the performance of the dynamic point cloud integration mechanism and confirmed that it can dynamically register the collected point clouds and update UDTs using a city-scale dataset.

However, the current implementation does not yet address aspects such as the quality of the collected point cloud data or the potential benefits of re-collecting data in previously scanned areas. Therefore, further analysis and improvements of the proposed components are necessary in future work.

Moreover, we plan to explore more advanced AR space representations that take into account the spatiotemporal degradation of point cloud quality. We will also investigate methods to guide participants toward spatially important regions, such as by utilizing the age of information. Additionally, we intend to conduct long-term experiments to comprehensively evaluate the overall operation and effectiveness of the proposed framework.

\section*{Acknowledgments}

This work was partly supported by NICT, Japan (JPJ012368C05601).


\end{document}